\documentclass[12pt,a4paper,twoside,frenchb]{article}
     \usepackage[latin1]{inputenc}
     \usepackage[T1]{fontenc}
     \usepackage[usenames,dvipsnames]{color}
     \usepackage{graphicx}
     \usepackage{babel}
     \usepackage{amssymb}
     \usepackage{verbatim}
     \usepackage[nomarkers]{endfloat}
     \usepackage{epsf,rotate,a4wide}
\newcommand{\lp}{\left(}
\newcommand{\rp}{\right)}
\newcommand{\lc}{\left[}
\newcommand{\rc}{\right]}

\newcommand{\dr}{\partial}
\newcommand{\pca}{{\cal P}}

\newcommand{\clst}{{\cal L}^*}

\newcommand{\cxst}{{\cal X}^*}
\newcommand{\cxba}{\overline{\cal X}}

\newcommand{\nd}{{\sf d}}

\newcommand{\Real}{{\rm I \! \! \! \; R}}

\newcommand{\tba}{\overline{T}}

\newcommand{\gammaba}{\overline{\gamma}}
\newcommand{\gammast}{\gamma^*}

\begin{document}

\newlength{\figwidth}
\setlength{\figwidth}{0.9\textwidth}

%

\vspace*{0.5cm}
\normalsize

\centerline{\bf On the Use of a Wider Class of Linear Systems for the Design of}
\medskip
\centerline{\bf Constant-Coefficients Semi-Implicit Time-Schemes in NWP}

\bigskip
\bigskip

\rm

\centerline{\sc P. B{\'e}nard$$ }
\bigskip
\bigskip
\centerline{\footnotesize \sl Centre National de Recherches Météorologiques, Météo-France, Toulouse, France}

\bigskip
\bigskip

\normalsize
\rm
\rm
\vspace{1in}
\centerline{12 September 2003}
\vspace{1in}
Corresponding address:

\medskip

Pierre Bénard

CNRM/GMAP

42, Avenue G. Coriolis

F-31057 TOULOUSE CEDEX

FRANCE

\bigskip

Telephone: +33 (0)5 61 07 84 63

Fax: +33 (0)5 61 07 84 53

e-mail: pierre.benard@meteo.fr

\newpage
\centerline {ABSTRACT}
\bigskip

The linearization of the meteorological equations 
around a specified reference state, usually applied
in NWP to define the linear system of constant-coefficients 
semi-implicit schemes, is outlined as an unnecessarily restrictive
approach which may be detrimental in terms of stability.
It is shown theoretically that an increased 
robustness can sometimes be obtained by
choosing the reference linear system in
a wider set of possibilities.
The potential benefits of this new approach 
are illustrated in two simple examples.
The advantage in robustness is not obtained 
at the price of an increased error or complexity.

\bigskip
\bigskip
\bigskip

\newpage

\section{Introduction}

The semi-implicit (SI) technique was proposed in the 70's 
(Robert {\sl et al.}, 1972) as a suitable and efficient
method for solving numerically the partial differential 
equations used in meteorology. 
At this time, the SI technique was applied to 
hydrostatic primitive equations (HPE), and its
success in this context made it very popular in the 
field of numerical weather prediction (NWP).
The suitability of the SI technique for 
the fully elastic Euler equations (EE) was 
then advocated (Tanguay et al. 1990), with the aim of
combining the advantages of a system valid at any scale 
and an efficient time-discretization, as required
for NWP purposes.

The essence of SI schemes is a linear separation 
of the source terms of the complete system to be 
solved, with an implicit treatment of this linear 
part. For the purpose of this paper, three main
types of SI schemes can be distinguished.
The coefficients of the implictly-treated linear
terms can be: 
(i) constant in time and horizontally 
  homogeneous (Simmons and Temperton, 1997; 
  Bubnov{\'a} et al., 1995, 
  Caya and Laprise, 1999); 
(ii) constant in time only (Thomas et al., 1998;
Qian et al., 1998); and 
(iii) non-constant (Skamarock et al., 1997,
  Cullen et al., 1997).

%

This paper only considers SI schemes belonging to the class
(i), which are designed under "constant-coefficient SI schemes"
in the following.
However, it should be outlined that since only 
the separation  of thermal terms is considered, 
all results and conclusions 
extend identically to those SI schemes of type 
(ii) for which the reference temperature is 
horizontally homogeneous (e.g. Thomas et al., 1998;
Qian et al., 1998).

The underlying principles usually applied in the 
design of constant-coefficients SI schemes are the following:

\begin{list}{}{}
\item[(i)] define a stationary SI reference basic state  $\cxst$;
\item[(ii)] linearize the meteorological system ${\cal M}$ to 
          be solved around this steady state,  to obtain a linear 
          system $\clst$;
\item[(iii)] treat the linear part of the evolution $\clst$
         with a centred-implicit scheme, and the remaining 
         "non-linear" part $({\cal M} - \clst)$ with a 
         centred-explicit scheme.
\end{list}

However, due to the explicit treatment of 
the non-linear (NL) residuals, the stability of this 
type of scheme is not formally guaranteed, especially with 
long time-steps.
Indeed, the application of the above technique sometimes 
leads to unexpected unstable behaviours. The two following 
problems (referred to as P1 and P2 hereafter) illustrate the 
kind of limitations which can be encountered with 
constant-coefficients SI schemes designed using  
the principles (i)--(iii):

\begin{list}{}{}
\item[P1:] With HPE, the introduction of a vertically varying 
 reference thermal profile $T^*$ close to the actual 
 atmospheric profile, although reducing 
 the magnitude of the thermal NL residuals, leads to a 
 scheme which is less robust than when a warm isothermal 
 $T^*$ profile is used (see e.g. Simmons et al. 1978, SHB78 
  hereafter).
\item[P2:] For two time-levels (2-TL) SI discretizations, the 
  EE system is extremely unstable while the HPE system
  is stable, as discussed in B{\'e}nard (2003, B03 hereafter). 
\end{list}

As mentioned above,
the constant-coefficients SI technique
has traditionally been applied to NWP by explicitly following 
the three principles (i)--(iii), but this method is unnecessarily
restrictive.

As stated in B03, the SI scheme can be viewed
as the very first iteration of a generalized 
pre-conditioned fixed-point
algorithm for iteratively approaching the pure 
centred-implicit scheme.
In this light, $\clst$ appears to be nothing
else than the linear pre-conditioner of the fixed-point 
algorithm (this pre-conditioner is necessary in such 
an algorithm for allowing the convergence of the 
iterative process).
This point of view outlines the arbitrariness of the 
choice of the $\clst$ system, provided 
a satisfactory convergence for the iterative 
algorithm is ensured.

When facing unexpected problems as (P1)--(P2), 
a possible solution, advocated in this paper, is
to relax the constraints (i)-(ii) and
to seek $\clst$ deliberately as an arbitrary 
constant-coefficients linear system, i.e. not obtained through the 
linearization of ${\cal M}$ around any 
reference state.
This method is illustrated 
in the two following practical examples.

\vspace{-0.4in}

\section{Proposed solution to the problem (P1)}
\label{sec_P1}

It is a well-documented fact that if the stability of 
the SI scheme is obtained by forcing NL residuals 
to large values (in such a way that their sign is 
controlled), then the response of the scheme is 
deteriorated, especially from the point of view
of phase-speed errors. When exaggerated, this 
strategy has a negative impact even on slower 
transient processes, making it unattractive
for NWP.
A natural way to alleviate this risk with 
certainty is thus to reduce the magnitude of 
NL residuals. This is precisely 
the idea which was tested in SHB78, by comparing
the properties of SI schemes obtained when 
choosing isothermal and non-isothermal 
profiles of the reference temperature $T^*$. 
The non-isothermal profiles were chosen 
close to standard atmospheric profiles, in 
such a way that the magnitude of thermal NL 
residuals was reduced compared 
to the case with isothermal $T^*$.
However, the experimental results in terms of 
stability were clearly worse for the 
non-isothermal option: when the tropopause of 
the $T^*$ profile was above the actual one, 
the scheme became highly unstable.
Given this experimental fact, the recommended 
solution, widely followed afterwards,
was to use a warm isothermal profile for $T^*$, 
thus implicitly accepting to sacrifice a better 
response of the scheme for an increased robustness.

\subsection{Analysis of SHB78 situation}
\label{sec_anal}

In this section, the HPE system in $\sigma$ coordinate
is considered with a three time-level (3-TL) leap-frog SI 
time-discretisation. 
The theoretical framework proposed in B03 is used
here to perform a stability analysis, and the reader is referred to this paper for
more details on notations and algebraic developments.
The framework is idealized in order to allow simpler 
analyses (Cartesian vertical $(x,z)$ plane without 
orography; dry, adiabatic, non-rotating equations).
A resting state $\cxba$ with a thermal profile 
$\tba(\sigma)$ is considered. 
All atmospheric evolutions are assumed to consist 
in small perturbations around $\cxba$ (referred to 
as the "actual" state hereafter), and the meteorological 
system ${\cal M}$ is linearized around this state, 
in order to allow tractable analyses.
In the notations of B03, the system thus writes:

\begin{eqnarray}
\frac{\dr D}{\dr t} & = & - R {\cal G} \frac{\dr^2 T}{\dr x^2}  
   - R \tba \frac{\dr^2 q}{\dr x^2}  
\label{eq_hpeD} \\
\frac{\dr T}{\dr t} & = & - \frac{R \tba}{C_p} {\cal S} D  
       - \lp \sigma \frac{d \tba}{d \sigma} \rp({\cal N}-{\cal S})D
\label{eq_hpeT} \\
\frac{\dr q}{\dr t} & = & -  {\cal N} D  
\label{eq_hpeq} 
\end{eqnarray}

\noindent where $D$ is the horizontal wind divergence, 
$T$ the perturbation temperature, $q=\ln(\pi_s)$ , and 
$\pi_s$ is the surface pressure. 
Note that in (\ref{eq_hpeT}), the last RHS term 
is the contribution of vertical advections for $\tba$.
This Eulerian form can be shown, in this linear framework,
to be also valid for a semi-Lagrangian discretization, 
under the assumptions of perfect solution for the 
displacement equation and perfect interpolators,
consistently with the current space-continuous context.
A modified version of the $\sigma$-coordinate static-stability 
for the actual state $\cxba$ is introduced through:

\begin{equation}
\gammaba = \frac{R \tba}{C_p} - \sigma  \frac{d \tba}{d \sigma}.
\label{eq_sts}
\end{equation}

\noindent
For the reference state $\cxst$ used to define the SI scheme, 
a profile $T^*(\sigma)$ is also assumed, 
and the system is linearized around this reference state,
according to the above principles (i)--(ii). 
The $\clst$ system and the static-stability $\gammast$ obtained 
through this procedure are thus formally 
identical to (\ref{eq_hpeD})--(\ref{eq_hpeq}) 
and (\ref{eq_sts}), respectively, simply substituting 
$T^*$ for $\tba$ everywhere. 

These two static-stabilities ($\gammaba$, $\gammast$) 
are now assumed uniform in the whole domain for the purposes
of the analysis. A "non-linearity" factor
is defined by $\alpha = (\gammaba - \gammast)/\gammast$.
It should be noted that the case of an isothermal SI 
reference state is also included in this formalism 
since it results in a uniform static-stability $\gammast$.
%
%
Following exactly the method presented in 
B03, the system (\ref{eq_hpeD})--(\ref{eq_hpeq}) is first 
transformed into an unbounded system:

\begin{eqnarray}
\lp \sigma \frac{\dr}{\dr \sigma}\rp  \frac{\dr D}{\dr t} & = &  R  \nabla^2 T 
\label{eq_hpeDu} \\
\lp {\cal I} + \sigma \frac{\dr}{\dr \sigma} \rp \frac{\dr T}{\dr t} & = & - \gammaba D  
\label{eq_hpeTu} \\
\end{eqnarray}

\noindent The normal modes of the system are then:

\begin{eqnarray}
\psi(x,\sigma) & = & \widehat{\psi} \; \exp(ikx) \sigma^{(i \nu - 1/2)} \\
\end{eqnarray}

\noindent where $(k,\nu) \in \Real$ and $\psi$ represents either $D$ or $T$.
Pursuing the analysis as in B03, it is finally found that
in the limit of long time-steps, the 3-TL SI scheme is 
stable for:

\begin{equation}
0 \leq \gammaba \leq 2 \gammast.
\label{eq_crit}
\end{equation}

\noindent This result extends (and is fully consistent with) those 
obtained in previous related studies 
(SHB78, and  C\^ot\'e et al. 1983, CBS83 hereafter).
Moreover, it allows an understanding of the instability 
observed in SHB78 for their SI scheme with non-isothermal
reference profiles: when the tropopause of the SI reference
state is higher than the tropopause of the actual state, 
the above criterion (\ref{eq_crit}) is locally violated between 
the two tropopauses, resulting in an unstable scheme.
However, for warm isothermal profiles of $T^*$, the 
latter instability disappears, as empirically found 
by SHB78, because in this case, $\gammast$ has a high
value at any level, and therefore is larger 
than $\gammaba/2$ in all the depth of atmosphere,
whatever may be the location of the actual tropopause.

\subsection{Proposed modification}

The fundamental difference between the two options
examined by SHB78 is not in the 
values of $T^*$ themselves (which actually deviate 
marginally between the two types of considered 
reference thermal profiles $T^*$), but in the presence 
or not of the advective term $(dT^*/d\sigma)$
in the $\clst$ system, because this term dramatically 
modifies the apparent static-stability $\gammast$,
as seen in (\ref{eq_sts}).

Hence, according to the new approach proposed in this paper,
a natural solution to ensure a more stable scheme 
while keeping a non-isothermal $T^*$ profile, 
is to deliberately remove the resulting advective 
term in the initial $\clst$ system.
This modification can be expected to combine
both the advantages of small residuals (because 
$T^*$ can be made closer to actual atmospheric
thermal profiles) and optimum stability 
(because the apparent static-stability in 
the $\clst$ system is large at any level).
It is worth noting that the mathematical structure 
of the $\clst$ system in this modified SI scheme 
with a non-isothermal $T^*$ profile 
is exactly the same as for a traditional SI scheme with 
an isothermal $T^*$, hence the modification in any 
pre-existing application is straightforward.


In order to illustrate the  consequences of this 
modification, a situation close to the one examined 
in SHB78 is considered. A class of vertical thermal 
profiles is introduced by :

\begin{equation}
T(\sigma)  =  {\rm max} \lc T_0, \lp T_0 - \frac{\gamma_0 C_p}{R} \rp 
     \lp\frac{\sigma}{\sigma_T} \rp^{(R/C_p)} + \lp \frac{\gamma_0 C_p}{R} \rp \rc,
\label{eq_prof}
\end{equation}

\noindent where $T_0=220$~K, $\gamma_0 = 30$~K, and
$\sigma_T$ is a varying parameter specifying the level 
of the tropopause. 
The value $\sigma_T^* = 0.25$ is chosen for the SI reference 
state $T^*$, while for the actual state $\tba$, the 
tropopause level $\overline{\sigma_T}$
is left as a free parameter in the interval $[0.1, 0.5]$. 
The static-stability is 
$\gamma = (R/C_p) T_0 = 62.9$~K in the isothermal 
$(T=T_0)$ "stratosphere", and $\gamma_0=30$~K
in the "troposphere" for both $\tba$ and $T^*$ 
profiles. The only difference between the two 
piecewise-constant profiles of static-stability 
($\gammaba$, $\gammast$) is thus the location of their 
tropopause.

The stability analysis is not straightforward for such 
multi-layers systems, hence the stability of the systems is
diagnosed through vertically-discretized analyses exactly
as in CBS83. 
In this method, the whole vertically- and time-discretized system 
for a given horizontal mode is considered 
as a linear "amplification matrix" acting on
a generalized vertical state-vector, and the growth rate 
$\Gamma$ of the system is the maximum modulus of the 
set of eigenvalues of the amplification matrix. 
The vertical structure of the
most-unstable mode is given by the associated
complex eigenvector.
The vertical discretisation in the analyses presented 
here is the same as in Simmons and Burridge (1981), and
is equivalent to the one used in SHB78.
The vertical domain is described through 80 
regularly-spaced $\sigma$ levels, and
the analyses are performed for a mode with 
$k=0.0005 {\rm m}^{-1}$ (the results are not
qualitatively sensitive to $k$).
As dicussed in B03, the examination of the stability
in the limit of long time-steps is relevant since
long time-steps area target in NWP.
The value chosen here is $\Delta t= 2000$~s
(here also, smaller time-steps do not change 
qualitatively the conclusions).

\begin{figure}[p]
\epsfxsize=\figwidth
\centerline{\epsfbox{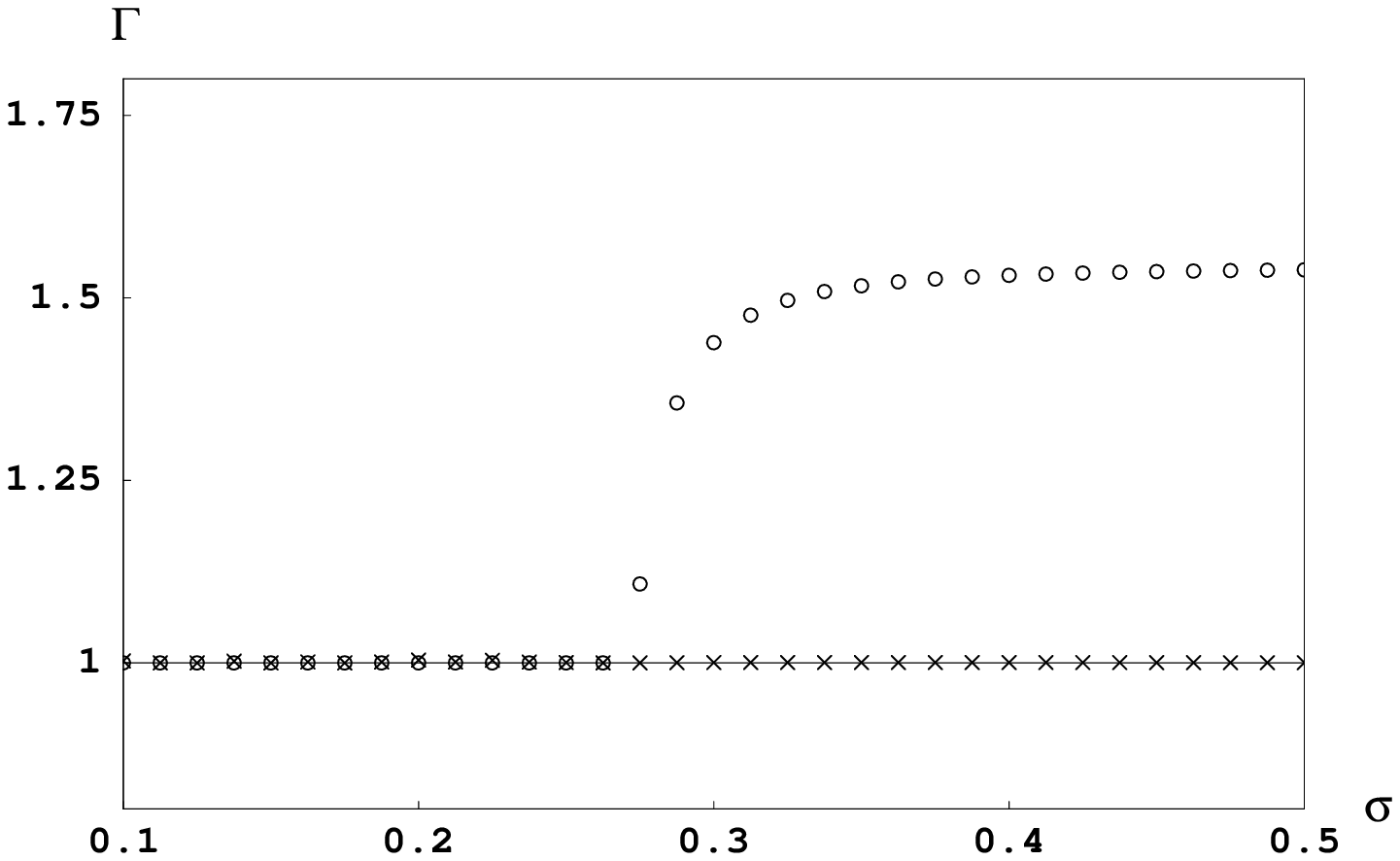}}
\caption{\label{fig_hpe} Growth-rate  $\Gamma$
 with the HPE system as a function of the 
 the tropopause location $\sigma$ for the thermal 
 profiles and discretisation settings as indicated 
 in the text.
  Circles: traditional 3-TL SI scheme; 
  crosses: modified 3-TL SI scheme.
  }
\end{figure}

The growth-rates for the traditional SI scheme 
and for the proposed modified SI scheme are depicted 
as a function of $\overline{\sigma_T}$ in Fig. \ref{fig_hpe}.
For the traditional SI scheme, the results are fully 
consistent with the criterion obtained through 
the above analysis  (it has also been
checked that a slight increase of the tropospheric 
static-stability $\gamma_0$ from 30 to 35~K results in a stable
scheme for any value of $\overline{\sigma_T}$ in 
the explored interval, in agreement with the 
stability criterion derived above).
For the traditional design used in SHB78, the
SI scheme is unstable as soon as the actual tropopause
is lower than its SI reference counterpart.
For the modified SI scheme proposed here, the 
stability is obtained even in the 
previously unstable situation, and
is thus comparable to the case with an isothermal 
$T^*$ profile.

In this analytical context, the modified 
scheme reaches the initial aim of reducing the 
magnitude of NL residuals while ensuring a robust
scheme.
The extension of these theoretical results to 
fully realistic frameworks has not been 
investigated further, but the approach 
seems worth considering since it potentially 
combines the two advantages of robustness and 
accuracy.

\section{Proposed solution to the problem (P2)}
\label{sec_P2}

A stability analysis of the EE system in the space-continuous 
SI framework, proposed in B03, shows that the 2-TL 
time-discretisation is very unstable in the presence of
thermal NL residuals, while the HPE system are acceptably
stable in the same context. 

As in the above section, a theoretical analysis
reveals the causes of this dramatic 
destabilization in simplified contexts.
In the following, it is shown that the destabilization
originates from the fact that the thermal NL 
residuals corresponding to the terms responsible for gravity 
and elastic waves systematically have opposite signs.
To better illustrate this explanation, the following couple 
of excerpts from the complete linearized EE system 
in $\sigma$ vertical coordinate [see (52)--(56) in B03] can be 
examined:

\begin{eqnarray}
\frac{\dr D}{\dr t} & = &  -  R {\cal G} \nabla^2 T 
\label{eq_exba}\\
\frac{\dr T}{\dr t} & = & - \frac{R \tba}{C_v} D
\label{eq_exbb}
\end{eqnarray} 

\noindent and:

\begin{eqnarray}
\frac{\dr \nd}{\dr t} & = & - \frac{g^2}{R \tba} 
  \lp \sigma \frac{\dr}{\dr \sigma} \rp \lp \sigma \frac{\dr}{\dr \sigma} +1 \rp\pca
\label{eq_exaa}\\
\frac{\dr \pca}{\dr t} & = & - \frac{C_p}{C_v} \nd
\label{eq_exab}
\end{eqnarray}

\noindent where all notations follow B03.

\noindent The first sub-system describes the horizontal propagation of
gravity waves, while the second describes the vertical propagation of
elastic waves. 
Neglecting the Boussinesq effect represented by the term "+1" 
in the RHS of (\ref{eq_exaa}), 
these two sub-systems are formally identical,
the only noticeable difference being the location of the $\tba$ factors
(at numerator vs. denominator). As a consequence, for a given set of actual and reference
temperatures ($\tba$, $T^*$) the thermal NL residuals always have an opposite 
sign in the two systems. 
For the purposes of the analysis, the thermal profiles $\tba$ and
$T^*$ can be considered as isothermal.
Let $\alpha=(\tba - T^*)/T^*$ be the thermal non-linearity 
parameter for the considered simplified problem.
The stability properties of the first sub-system for $\alpha$ are thus
the same as those of the second sub-system for $-\alpha/(1+ \alpha)$. 
Since the stability of the first sub-system (in 2-TL SI) for long time-steps
implies $\alpha \leq 0$ (see B03 for details), the stability of 
the second sub-system necessarily implies $\alpha \geq 0$, and 
the stability domain for a complete SI system which would 
include the two types of waves thus vanishes.
In other words, if $T^*$ is chosen so as to stabilize horizontally propagating
gravity waves, then vertically propagating elastic waves will be unstable, 
and {\sl vice versa}.
The problem is of course not present for HPE since this system does
not allow the propagation of elastic waves.

A natural solution to restore systematically the same sign 
for thermal NL residuals in the above two sub-systems, is thus 
to introduce different values of $T^*$ for each sub-system,
that is: $T^*$ for the gravity-wave system 
(\ref{eq_exba})--(\ref{eq_exbb}), and $T^*_E$ for
the elastic-wave system (\ref{eq_exaa})--(\ref{eq_exab}).
Noting $T^*_E = r T^*$, the stability domains for the first 
and second system become $\alpha \leq 0$ and  $(r-1) \leq \alpha$
respectively. As a consequence, choosing $r<1$ allows  
a non-empty stability domain for long time-steps to be restored.
In terms of temperature, the stability is then ensured if
$T^*_E \leq  \tba \leq  T^*$ in this isothermal context.
The stability domain for $\tba$ 
can thus be arbitrarily extended, by setting $T^*$ arbitrarily 
warm, and $T^*_E$ arbitrarily cold, with the 
limitation that an exaggeration in this direction finally
deteriorates the response of the scheme, as outlined above.

The application of  this solution to the complete EE system is 
straightforward: for all occurences of $T^*$ at numerator
in the initial linear system [i.e. the reference system 
$\clst$ associated to (52)--(56) in B03], the traditional
warm value  $T^*$ should be kept, while for the occurences of $T^*$ 
at the denominator, the cold value $T^*_E$ should be imposed.
Here also, the modification from any pre-existing
application is straightforward.

\begin{figure}[p]
\epsfxsize=\figwidth
\centerline{\epsfbox{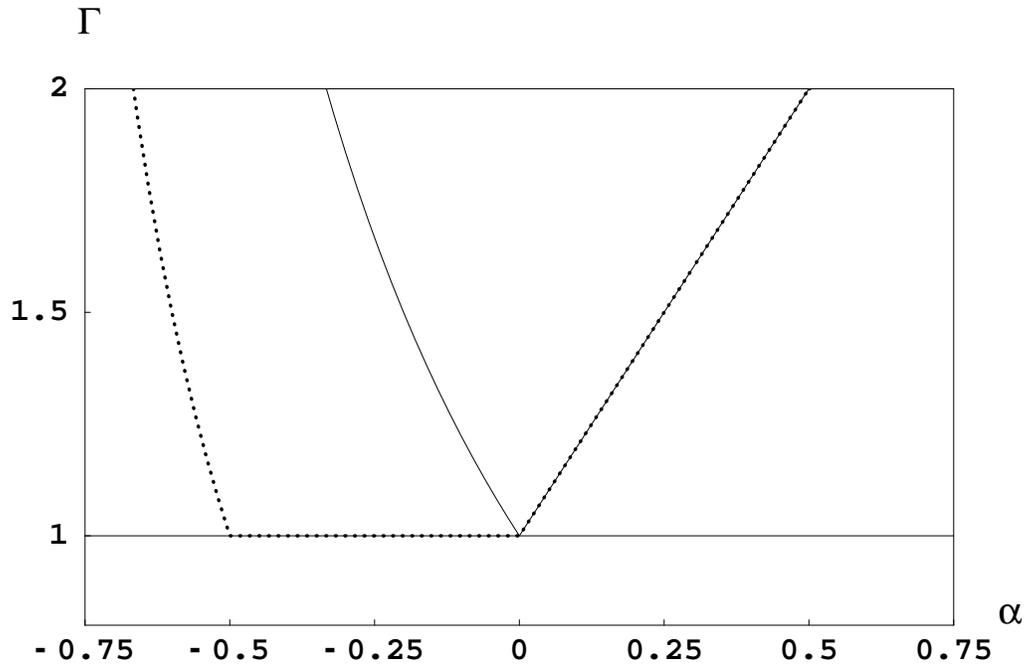}}
\caption{\label{fig_ee1} Analytical growth-rate $\Gamma$
 for the EE system as a function of $\alpha$ 
 in the limit of large time-steps.
  Solid line: traditional 2-TL SI scheme; 
  dotted line: modified 2-TL SI scheme.
  }
\end{figure}

The theoretical impact of this modification 
is first illustrated with a stability analysis of the
complete EE system for 2-TL SI schemes in the
context of isothermal $\tba$ and $T^*$ profiles
and linear evolution around $\cxba$ as in B03. The analysis
for $T^*_E=T^*$ is given in B03, and can be repeated
in a formally similar way for the modified SI scheme 
$T^*_E \neq T^*$. The growth rates
obtained in the long time-step limit for the initial 
and modified SI schemes are depicted in Fig. \ref{fig_ee1},
for $r=1$ and $r=0.5$ (i.e. $T^*_E = T^* /2$).
The results are fully consistent with the above 
simple analyses: the modified SI scheme is found to be stable in 
the interval $(r-1) \leq \alpha \leq 0$, while the traditional
SI scheme is always unstable.

In order to evaluate the potential benefit of the proposed 
approach for NWP, the modification was then tested in real-case
conditions with the adiabatic semi-Lagrangian version of the Aladin-NH model 
(Bubnov{\'a} et al. 1995), used with a 2-TL SI time-discretization. 
The model was integrated for 3 hours 
for a randomly-chosen situation consisting of a strong flow
over real topography, in a domain which includes the 
montanous Pyr{\'e}n{\'e}es region. The horizonal resolution
is 2.5 km in horizontal directions, and the time-step is 80~s. 
The vertical coordinate is the mass-based hybrid coordinate 
defined in Simmons and Burridge (1981), and the domain
is discretised along 41 irregular layers with a thickness
increasing with height, in the usual NWP fashion. 
Integrations are performed without 
any time-filter (see B03 for a discussion on the detrimental
effects of  time-filters in 2-TL SI EE system).
A weak fourth-order horizontal diffusion is applied to 
avoid the accumulation of energy in the smallest resolved 
scales during the course of the integration.

\begin{figure}[p]
\epsfxsize=\figwidth
\centerline{\epsfbox{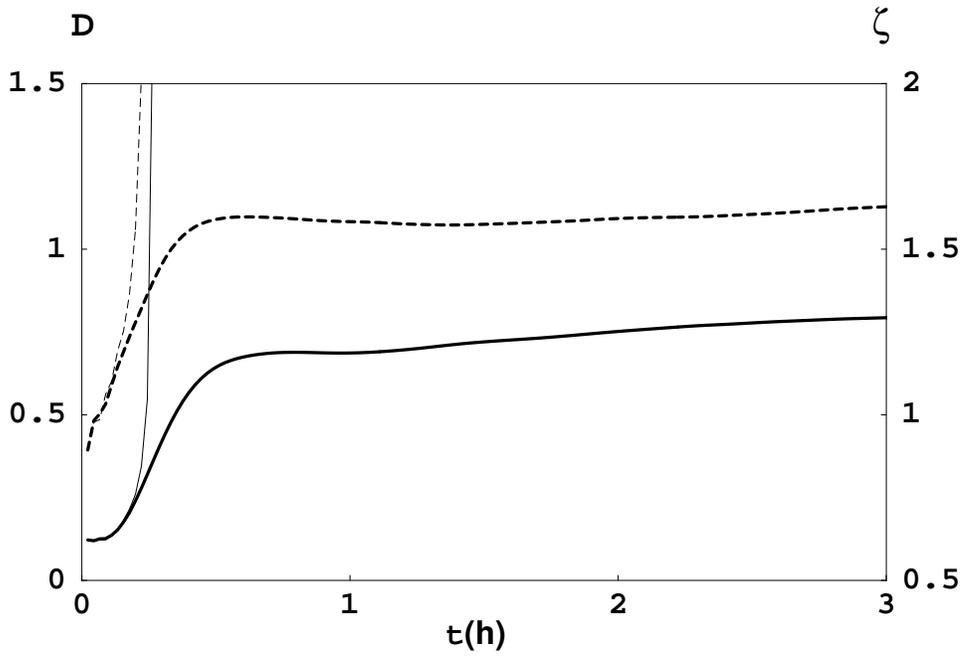}}
\caption{\label{fig_ee2} Evolution of the spectral norm of
vorticity $\zeta$ and horizontal divergence $D$ for a real-case
with a 2-TL SI EE system.
  Solid line: vorticity (right axis); 
  dashed line: divergence (left axis);
  thin line: traditional 2-TL SI scheme; 
  thick line: modified 2-TL SI scheme.
  }
\end{figure}

Fig. \ref{fig_ee2} shows the evolution of the whole domain 
spectral norms of the horizontal vorticity $\zeta$ and 
divergence $D$ for 
the traditional and modified versions of the 2-TL SI scheme.
The traditional SI scheme is used 
with $T^* = T_E^*$=300K, and the modified SI scheme
with $T^*$=300~K, $T_E^*$=150~K.
The original 2-TL SI scheme is clearly unstable, since the
integrations diverge after 11 time-steps, while the modified 
2-TL SI scheme behaves stably during the 3 hours of the 
integration. This experiment clearly indicates a potential
advantage of using  the modified SI scheme in NWP
with 2-TL EE systems.

\section{Comments and conclusion}
\label{sec_conc}

All the discussions in this paper apply equally to 2-TL and 3-TL
schemes. They can also be extended straightforwardly from SI schemes to 
the emerging class of iterated centred-implicit (ICI) schemes,
as examined in B03, because these schemes are based 
on the same kind of linear separation of the meteorological 
system to be solved implicitly.

For the problem (P2), 
the proposed solution offers a smaller
interest for 3-TL schemes than for 2-TL schemes, 
because 3-TL schemes already have a
degree of robustness compatible with a 
NWP use as far as thermal NL residuals are 
concerned. 
Nevertheless, the proposed solution is believed 
to be worth considering for high resolution modelling 
with the EE system in combination with 2-TL schemes
For the EE system discretized with
2-TL SI schemes, it may allow in particular to remove 
the strong time-filters used so far, with their detrimental 
effects in terms of response.
Moreover, it would 
be interesting to extend this modification to systems in 
height-based coordinates. It is worth noting also that the two
modifications proposed in sections \ref{sec_P1}
and \ref{sec_P2} could be
combined to obtain a stable 2-TL scheme together with 
thermal NL residuals of smaller magnitude.

More generally, the aim of this paper is to emphasize that
there may be a considerable benefit to relax the unnecessarily 
constraining principles (i)--(ii) for the design of 
all kinds of implicit schemes
based on a linear separation (SI and ICI schemes).
In practice, starting from an initially unstable 
scheme obtained through the traditional approach,
a dramatic improvement may sometimes be obtained if
the approach proposed here is used for modifying
this initial scheme in only slight details. 
The discussions in this paper clearly do not offer 
a complete picture of the properties
of the modified schemes proposed above compared 
to their traditional counterparts.
Before extending such modifications to the actual NWP framework,
it would be necessary to evaluate more precisely their practical impact 
in terms of accuracy and response corectness on forecasts
performed in real conditions.


\newpage

\section*{References}

\begin{description}

\item Bénard, P., 2003:
      Stability of Semi-Implicit and Iterative Centered-Implicit Time
      Discretizations for Various Equation Systems Used in NWP.
      {\em Mon. Wea. Rev.}, {\bf 131}, 2479-2491.
      
\item Bubnov\'a, R., G. Hello, P. B\'enard, and J.F. Geleyn, 1995:
      Integration of the fully elastic equations cast in the hydrostatic
      pressure terrain-following coordinate in the framework of the
      ARPEGE/Aladin NWP system.
      {\em Mon. Wea. Rev.}, {\bf 123}, 515-535.

\item C\^{o}t\'{e}, J., M. B\'eland, and A. Staniforth, 1983:
      Stability of vertical discretization schemes for semi-implicit
      primitive equation models: theory and application.
      {\em Mon. Wea. Rev.}, {\bf 111}, 1189-1207.

\item Cullen, M. J. P., T. Davies, M. H. Mawson, J. A. James 
      and S. C. Coulter, 1997: An overview of numerical methods 
      for the next generation UK NWP and climate model.
      {\em Numerical Methods in Atmospheric Modelling}, 
      Canadian Meteorological and Oceanographical Society,
      581 pp.

\item Qian, J.-H., F. H. M. Semazzi, and J. S. Scroggs, 1998:
      A global nonhydrostatic semi-Lagrangian atmospheric model with
      orography.
      {\em Mon. Wea. Rev.}, {\bf 126}, 747-771.

\item Robert, A. J., J. Henderson, and C. Turnbull, 1972: An implicit time
      integration scheme for baroclinic models of the atmosphere . 
      {\em Mon. Wea. Rev.}, {\bf 100}, 329-335.

\item Simmons, A. J., B. Hoskins, and D. Burridge, 1978:
      Stability of the semi-implicit method of time integration.
      {\em Mon. Wea. Rev.}, {\bf 106}, 405-412.

\item Simmons, A. J., and D. Burridge, 1981:
      An Energy and Angular-Momentum Conserving Vertical Finite-Difference
      Scheme and Hybrid Vertical Coordinates.
      {\em Mon. Wea. Rev.}, {\bf 109}, 758-766.

\item Simmons, A. J., C. Temperton, 1997:
      Stability of a two-time-level semi-implicit integration scheme
      for gravity wave motion.
      {\em Mon. Wea. Rev.}, {\bf 125}, 600-615. 
      
\item Skamarock, W. C., P. K. Smolarkiewicz, and J. B. Klemp, 1997:
      Preconditioned conjugate-residual solvers for Helmholtz equations
      in nonhydrostatic models.
      {\em Mon. Wea. Rev.}, {\bf 125}, 587-599. 

\item Tanguay, M., A. Robert, and R. Laprise, 1990:
      A semi-implicit semi-Lagrangian fully compressible regional forecast model.
      {\em Mon. Wea. Rev.}, {\bf 118}, 1970-1980.

\item Thomas, S.J., C. Girard, R. Benoit, M. Desgagn{\'e}, and P. Pellerin, 1998:
      A new adiabatic kernel for the MC2 model.
      {\em Atmos. Ocean}, {\bf 36 (3)}, 241-270.

\end{description}

\newpage
\section*{List of Figures}

 Fig. 1:  Growth-rate  $\Gamma$
 with the HPE system as a function of the 
 the tropopause location $\sigma$ for the thermal 
 profiles and discretisation settings indicated 
 in the text.
  Circles: traditional 3-TL SI scheme; 
  crosses: modified 3-TL SI scheme.

\medskip

Fig. 2: Growth-rate $\Gamma$ as a function of $\alpha$,
 in the limit of long time-steps
 for the complete EE system examined in section \ref{sec_P2}.
  Solid line: traditional 2-TL SI scheme; 
  dotted line: modified 2-TL SI scheme.
  
\medskip

Fig. 3: Evolution of the spectral norm of
vorticity $\zeta$ and horizontal divergence $D$ for a real-case
with a 2-TL SI EE system.
  Solid line: vorticity (right axis); 
  dashed line: divergence (left axis);
  thin line: traditional 2-TL SI scheme; 
  thick line: modified 2-TL SI scheme.

\newpage



\end{document}